\documentclass[a4paper,aps,prl,twocolumn,showpacs]{revtex4}
\usepackage{amsmath}
\usepackage{hyperref}
\begin{document}
\title{Comment on ``Spin-Gradient-Driven Light Amplification in a Quantum Plasma''}
\author{Govind S. Krishnaswami$^{1}$, Rajaram Nityananda$^{2}$, Abhijit Sen$^{3}$, Anantanarayanan Thyagaraja$^{4}$} 
\affiliation{$^{1}$Chennai Mathematical Institute,  SIPCOT IT Park, Siruseri 603103, India\\
$^{2}$National Centre for Radio Astrophysics, TIFR, Pune 411007, India\\
$^{3}$Institute for Plasma Research, Bhat, Gandhinagar 382428, India\\
$^{4}$Astrophysics Group, University of Bristol, Bristol, BS8 1TL, United Kingdom\\
{\rm Email: govind@cmi.ac.in, rajaram@ncra.tifr.res.in, senabhijit@gmail.com, a.thyagaraja@bristol.ac.uk}}
\pacs{52.35.-g, 52.35.Hr; 52.35.We; 67.10.-j; 67.30.hj}



\begin{abstract}

A comment on the Letter by S. Braun, F. A. Asenjo and S. M. Mahajan, Phys. Rev. Lett., {\bf 109}, 175003 (2012). We show that recent arguments for light amplification driven by inhomogeneous quantum spin fields in low temperature electron plasmas in metals are invalid. In essence, a neglect of Pauli `blocking' led the authors to over-estimate the effects of intrinsic spin. \\

\centerline{Journal Reference: Phys. Rev. Lett. 112, 129501 (2014). Submitted 17 Aug, 2013, accepted 19 Feb, 2014.}

\end{abstract}
\maketitle

Recently, Braun, Asenjo, and Mahajan \cite{BraunAsenjoMahajan} made a startling prediction: an electromagnetic (EM) wave above the cut off frequency, incident on a metal at low temperatures ($\sim 30$ K), will amplify in the presence of a conduction electron spin field ${\bf S}$ with large gradients. This prediction is based on spin quantum hydrodynamic (SQHD) equations derived in \cite{Marklund} and the `quantum spin vorticity' formalism of \cite{MahajanAsenjo}. The authors treat the metal as a free electron plasma with a uniform neutralizing ion background. However, their equations are not valid for conduction electron densities ($n \simeq 10^{29}/$m$^3$) in metals at low temperatures, due essentially to a neglect of Pauli blocking in the derivations of \cite{Marklund, MahajanAsenjo}. For such $n$, the Fermi temperature $T_{\rm F} = ({\hbar^2/2 m_e}) (3\pi^2 n)^{2/3}$ is a few eV. Hence, even at room temperature, the electron gas is highly degenerate. When $T \ll T_{\rm F}$ the (thermal) de Broglie wavelength exceeds the interelectron distance,  the electron gas obeys Fermi-Dirac (FD) statistics. Unfortunately, the authors' equations (based on \cite{Marklund}) assume a factorized form for the $N$-electron wave function, ignoring the antisymmetrization required by the Pauli exclusion principle. Consequently, the electron spin magnetization current is greatly overestimated. They use the formula ${\bf j}_{\rm sp} = \nabla \times [2n\mu_{B}{\bf S}]$ [Eqs. (2), (3) in \cite{BraunAsenjoMahajan}] where $\mu_B = e \hbar/2m_e$. However, in a degenerate electron gas, the spin magnetization ought, by Pauli blocking, to be of order $2(n_+ - n_-) \mu_B {\bf b}$ where ${\bf b}$ is the unit vector in the local magnetic field (${\bf B}_0$) direction. Here, $n_\pm = \frac{n}{2} \left[ 1 \pm \frac{3 \mu_B B_0}{2 T_F} \right]$ \cite{LifshitzPitaevski,AshcroftMermin} are the number densities of electrons with spins parallel and antiparallel to ${\bf B}_0$ and ${(n_+ - n_-)}/{n} = {\cal O}(T/T_F)$. At temperatures $T \ll T_F$, only electrons close to the Fermi level [estimated by $(T/T_F)n$] contribute to the magnetization current, as well as any electron dynamics. Thus, $\bf S$ must have a size $\simeq [n_+ - n_-]{\bf b}/n $, invalidating the authors' assumption [Eq. (1) of \cite{BraunAsenjoMahajan}] that $\bf S$ is a {\em unit vector}. This assumption leads to a huge $j_{\rm sp} \simeq  2n({\mu_B}/{L_{S}}) \simeq 10^{6}{\rm A.m}^{-2}$, where, $L_S = 1$m is a macroscopic gradient length-scale. Any smaller $L_S$ will make $j_{\rm sp}$ even larger. This corresponds to a near `saturation magnetic induction' of 1 Tesla, a value more typical of spin-polarized core electrons in a ferromagnet than conduction electrons in a metal. A saturation magnetization is problematic, coming at the cost of Fermi energy, which corresponds to creating a highly excited state. Thus, when $T \ll T_F$, Pauli blocking ensures that quantum spin effects are suppressed, as is well known \cite{AshcroftMermin} from the smallness of Pauli paramagnetism and Landau diamagnetism. Fermi liquid theory and quantum Boltzmann equations are required in this regime\cite{LifshitzPitaevski}. 
On the other hand, when $T \gg T_F$ and Maxwell-Boltzmann statistics apply, quantum spin effects are negligible, being a small perturbation [of ${\cal O}(\mu_B B/T) \ll 1$] to the standard Vlasov kinetics (as also stated in \cite{Marklund}). At high temperatures, Larmor moments $\mu_L = T/B$ dominate over spin moments ($\mu_L \gg \mu_B$). Also, at any $T$, the quantum spin force $\mu_B \nabla B$ is a very small perturbation to the classical orbit theory based on the Lorentz force. Furthermore, Coulomb collisions are non-negligible at high densities. The authors require a low $T \simeq 30$K to prevent collisions from damping the EM wave moving into their inhomogeneous spin density-dominated medium. This low collision rate is due to Pauli blocking\cite{LifshitzPitaevski,AshcroftMermin}. It is {\em inconsistent} to use Pauli blocking to suppress collisions, but ignore its effect on the spin magnetization. Owing to these omissions and contradictions, the authors' equations are valid neither at high nor low $T$. Moreover, they do not consider the equation of state of the electron gas nor the basic equilibrium state [involving significant spin gradients], around which they linearize their equations [(6)-(8) of \cite{BraunAsenjoMahajan}] for spin and EM fields. Discussion of the conditions needed [in principle] to create such a medium, and a Poynting theorem describing the pumping of the EM wave by the spin gradients, would lend credence to their predictions. For other critiques of hydrodynamical treatments of quantum  plasmas (``QHD'') see \cite{Poedts1,Bonitz}. This Comment (see \cite{critique-arxiv} for a more detailed discussion and references) focuses on the erroneous treatment of electrons in \cite{BraunAsenjoMahajan}. 


\clearpage

\centerline{\bf Addendum to Comment}
\centerline{Date: March 2, 2014.}

\vspace{.1cm}

In \cite{mahajan-asenjo-reply}, the authors of \cite{BraunAsenjoMahajan} offer a Reply to our Comment. They accept the validity of our criticisms. In particular, they agree that (1) the average spin field $\bf S$ is not a unit vector and (2) that Pauli blocking will greatly reduce the spin magnetization current in low temperature metallic plasmas, and thereby reduce any instability. They go on to state that at low temperatures, the light wave growth rate as well as electron-electron collision/damping frequency will both be brought down by the factor $\alpha = T/T_F$, yielding
	\begin{equation}
	\Gamma_{\rm PB} = \alpha \Gamma_{old} \quad \text{and} \quad
	\nu_{ee} \sim \frac{k_B T^2}{\hbar T_F}.
	\end{equation}
Thus they estimate the ratio of collisional damping to growth rate as
	\begin{equation}
	\frac{\nu_{ee}}{\Gamma_{\rm PB}} \sim \frac{k_B T}{\hbar \Gamma_{old}}.
	\end{equation}
They assert that, in principle, this ratio could be less than unity for sufficiently low temperatures, thereby implying amplification of the light wave.

It is true that in an `ideal metal', the electron collision rate $\nu_{ee}$ will scale like $T^{2}$ [cf. \cite{LifshitzPitaevski}, also Eq.(17.66) in \cite{AshcroftMermin}]. In reality, at very low temperatures [when, $T \ll T_{\rm Debye}$], the `residual resistance' of a metal due to impurity or lattice defect scattering leads to a temperature-independent collision rate $\nu_0$. If such a temperature-independent collision frequency is used, then the authors' estimate for the ratio of collision rate to growth rate $\frac{\nu_0}{\Gamma_{\rm PB}} \sim \frac{T_F \nu_0}{T \Gamma_{\rm old}}$ would be  {\em more} than unity for sufficiently low temperatures, implying that collisions prevent any light amplification.

Even if we accept $\nu_{ee} = k_B T^2/\hbar T_F$ as a reasonable low temperature collision rate, as well as their method of estimating the effect of Pauli blocking [i.e., that $\Gamma_{\rm PB} = \frac{T}{T_F} \Gamma_{\rm old}$], we find for solid state plasmas, that the ratio of collision to growth rate $\nu_{ee}/\Gamma_{\rm PB}$ is less than unity only for very low temperatures $T \lesssim 0.025 K$. Collisional damping of the wave will overwhelm the claimed effect for any higher temperature. What is more, even at such a low temperature, the EM wave would have to travel at least $c/\Gamma_{PB} \sim 30$ km in their medium to be significantly amplified. To obtain these estimates for solid state plasmas considered by the authors, we take $n_e \approx 10^{29}$m$^{-3}$, corresponding to a Fermi temperature $k_B T_F \approx 1$ eV $\approx 10^4$ K and a plasma frequency $\omega_{pe} \approx 1.6 \times 10^{16}$ s$^{-1}$. Now from Fig. 1 of their Letter \cite{BraunAsenjoMahajan}, the {\em maximum} value of $\Gamma_{\rm old}$ is $2.5 \times 10^{-7} \times \omega_{pe} \approx 4 \times 10^{9}$ s$^{-1}$. Using their formulas $\nu_{ee} \sim \frac{k_B T^2}{\hbar T_F}$ and $\frac{\nu_{ee}}{\Gamma_{PB}} \sim \frac{k_B T}{\hbar \Gamma_{\rm old}}$, we find that $k_B T \lesssim 2.5 \times 10^{-6}$ eV $\approx 0.025$K for the growth rate to exceed the collision rate. At $T = 0.025$ K, $\Gamma_{PB} \approx 10^4$ s$^{-1}$ and so the wave must travel $c/\Gamma_{PB} \approx 30$ km in the medium to be amplified significantly. Thus, even assuming SQHD to be corrected as suggested by the authors, the effects predicted are negligible [using their own numbers and formulae] and are far smaller than many other neglected effects such as collisionless damping, impurity scattering, etc.

However, there is a more basic problem. The authors estimate the effect of FD statistics on the growth rate to be given by simply multiplying the old (uncorrected) growth rate $\Gamma_{\rm old}$ by $(\frac{T}{T_{F}})$. This is unacceptable, since the growth rate has not been derived ab initio using {\it correct} SQHD equations when $T \ll T_{F}$. The correct equations will have drastically reduced spin forces in the electron momentum equation and much lower spin magnetization currents in Maxwell's equations. It is to be checked {\em afresh} using correct equations, whether the mode will grow {\em at all}, for the stated conditions [especially the `WKB' condition $kL_{S} \gg 1$, where $k$ is the wave number and $L_{S}$ the gradient length-scale of the thermally averaged spin field].

As mentioned in our Comment, it is readily shown from a consideration of the single-particle electron Hamiltonian in classical physics, that an added `spin magnetic moment force' [$\simeq |\mu_{\rm Bohr}\nabla B|$] borrowed from relativistic quantum mechanics, is tiny compared to the Lorentz forces arising from self-consistent electromagnetic fields. The spin-dependent dipole force can be shown from perturbation theory, to modify standard results of wave propagation in plasmas by minuscule effects. When large numbers of electrons are considered, there is the further crucial point that the spins can be oriented [in quantum theory] along or anti-parallel to the local magnetic field, and there will be very large cancellations of an already small effect by quantum and thermal averaging. The authors offer no clear argument to evade these standard conclusions.


\begin{thebibliography}{99}
\bibitem{BraunAsenjoMahajan} S. Braun, F. A. Asenjo and S. M. Mahajan, Phys. Rev. Lett., {\bf 109}, 175003 (2012). 
\bibitem{Marklund} M. Marklund and G. Brodin, Phys. Rev. Lett., {\bf 98}, 025001 (2007).
\bibitem{MahajanAsenjo} S. M. Mahajan, F. A. Asenjo, Phys. Rev. Lett., {\bf 107}, 195003 (2011).
\bibitem{LifshitzPitaevski} E. M. Lifshitz and L. P. Pitaevski, {\em Physical Kinetics}, (Pergamon, Oxford, 1981) Chap. VIII.
\bibitem{AshcroftMermin} N. W. Ashcroft and N. D. Mermin, {\em Solid State Physics}, (Harcourt Brace, New York, 1976), Chaps. 2 and 31.
\bibitem{Poedts1} J. Vranjes, B. P. Pandey and S. Poedts, Europhys. Lett., {\bf 99}, 25 001 (2012).
\bibitem{Bonitz} M. Bonitz, E. Pehlke and T. Schoof, Phys. Rev. E, {\bf 87}, 033105 (2013).
\bibitem{critique-arxiv} G. S. Krishnaswami, R. Nityananda, A. Sen and A. Thyagaraja, {\em A critique of recent theories of spin half quantum plasmas},  arXiv:1306.1774 [physics.plasm-ph].
\bibitem{mahajan-asenjo-reply} S. Braun, F. A. Asenjo and S. M. Mahajan, Reply to Comment on ``Spin-Gradient-Driven Light Amplification in a Quantum Plasma'', Phys. Rev. Lett. {\bf 112}, 129502 (2014).

\end{thebibliography}
\end{document}